\documentclass[conference]{IEEEtran}
\IEEEoverridecommandlockouts
\usepackage{cite}
\usepackage{amsmath, amssymb, amsfonts}
\usepackage{algorithmic}
\usepackage{graphicx}
\usepackage{textcomp}
\usepackage{soul}

\usepackage{xcolor}
\usepackage{pifont}
\usepackage{subcaption}
\def\BibTeX{{\rm B\kern-.05em{\sc i\kern-.025em b}\kern-.08em
    T\kern-.1667em\lower.7ex\hbox{E}\kern-.125emX}}
\def\BibTeX{{\rm B\kern-.05em{\sc i\kern-.025em b}\kern-.08em
    T\kern-.1667em\lower.7ex\hbox{E}\kern-.125emX}}

\begin{document}

\title{Efficient IoT Intrusion Detection with an Improved Attention-Based CNN-BiLSTM Architecture\\
}

\author{
\IEEEauthorblockN{Amna Naeem}
\IEEEauthorblockA{
\textit{Department of Computer Science} \\
\textit{Quaid-i-Azam University}\\
Islamabad, Pakistan \\
amna@cs.qau.edu.pk}

\and
\IEEEauthorblockN{Jawad Ahmad}
\IEEEauthorblockA{
\textit{Cyber Security Center} \\
\textit{Prince Mohammad Bin Fahd University}\\
Alkhobar, Saudi Arabia \\
jahmad@pmu.edu.sa}

\and
\IEEEauthorblockN{Muazzam A. Khan}
\IEEEauthorblockA{
\textit{Department of Computer Sciences} \\
\textit{Quaid-e-Azam University}\\
Islamabad, Pakistan \\
muazzam.khattak@qau.edu.pk}

\and
\IEEEauthorblockN{\hspace*{\fill}Aizaz Ahmad Khattak\hspace*{\fill}}
\IEEEauthorblockA{
\textit{School of Computing, Engineering and Built Environment} \\
\textit{Edinburgh Napier University}\\
Edinburgh, UK \\
40614576@live.napier.ac.uk}

\and
\IEEEauthorblockN{\hspace*{\fill}Muhammad Shahbaz Khan\hspace*{\fill}}
\IEEEauthorblockA{
\textit{School of Computing, Engineering and Built Environment} \\
\textit{Edinburgh Napier University}\\
Edinburgh, UK \\
muhammadshahbaz.khan@napier.ac.uk}
}

\maketitle

\begin{abstract}
The ever-increasing security vulnerabilities in the Internet-of-Things (IoT) systems require improved threat detection approaches. This paper presents a compact and efficient approach to detect botnet attacks by employing an integrated approach that consists of traffic pattern analysis, temporal support learning, and focused feature extraction. The proposed attention-based model benefits from a hybrid CNN-BiLSTM architecture and achieves 99\% classification accuracy in detecting botnet attacks utilizing the N-BaIoT dataset, while maintaining high precision and recall across various scenarios. The proposed model's performance is further validated by key parameters, such as Mathews Correlation Coefficient and Cohen’s kappa Correlation Coefficient. The close-to-ideal results for these parameters demonstrate the proposed model's ability to detect botnet attacks accurately and efficiently in practical settings and on unseen data. The proposed model proved to be a powerful defense mechanism for IoT networks to face emerging security challenges.
\end{abstract}


\section{Introduction}

The application of Artificial Intelligence (AI) in Internet-of-Things (IoT) has resulted in innovative and  sophisticated networked systems. Such systems, comprising of a number of devices ranging from simple sensors to advanced machinery, influence a number of applications, e.g., smart cities, homes, transportation, manufacturing, and agriculture \cite{9520818}. IoT devices have become deeply embedded in everyday life. However, this extensive adoption brings significant security risks. Due to their diverse nature and inherent vulnerabilities, IoT devices are particularly prone to malicious activities or cyber attacks such as malware attacks. As the use of IoT devices continues to grow, concerns over security and privacy are rising, given the increasing exposure of sensitive data online. To reduce cyber attacks targeting IoT networks, continuous monitoring is crucial.  Analyzing IoT network traffic can enhance intrusion detection mechanisms within the IoT ecosystem,ultimately strengthening cybersecurity \cite{10471929}. Intrusion detection systems serve as essential supplementary security measures, operating in conjunction with other defenses to reduce potential threats. Deep Learning a subset of machine learning is capable of extracting relevant information from massive and high-speed network traffic generated by various interconnected IoT devices \cite{popoola2020hybrid}.The presented hybrid deep learning security framework integrates the attention mechanism with 1D-CNN and BiLSTM. The framework demonstrates robust intrusion detection capabilities using normalized data and measuring performance using the N-BaIoT dataset across common metrics. Common metrics are used to evaluate the performance of each model, taking into account the specific challenges of the data sets, including various device behaviors and class imbalance \cite{10836622} This study presents a few significant advancements which are as follows:
\begin{itemize}
    \item The proposed attention-based 1D-CNN + BiLSTM model efficiently and accurately detects botnet attacks while maintaining high-precision and recall in various scenarios. 
    \item The proposed model demonstrates a reduced computational overhead without compromising the detection accuracy.
    \item The proposed model has been compared with the latest deep learning approaches demonstrating its superior performance.
\end{itemize}

\section{Literature Review}
The growing number of limited resources IoT devices is becoming more harder to protect as novel security risks emerge. Researchers have been practicing with a number of machine learning methods utilizing N-BaIoT data to figureout these security concerns, for instance, authors in  \cite{9120761} demonstrated an IDS for IoT industry 4.0. They merged cooperative learning and Deep Neural Networks (DNN) approaches and the proposed scheme scored well, obtaining an 0.92\% as F1-Score and a 95.06\% of accuracy.  Another study \cite{10136065}, on several neural network structures, discovered that CNN, RNN and LSTM outperformed in botnet identification with an accuracy of 91\%, 41\% and 62\%, respectively. Furthermore, the researchers in \cite{majd2023iot} developed a CNN-based deep learning model for botnet detection using the N-BaIoT dataset, improving scalability in diverse environments. The model showed superior accuracy, precision, and recall, but struggled with UDP-based attacks. The research emphasizes the effectiveness of CNNs in IoT security. In addition, a novel CNN-LSTM model \cite{https://doi.org/10.1155/2021/3806459} was introduced to identify BASHLITE and Mirai security attacks, achieving an accuracy of 90.88\% and precision and recall rates of 93.04\% and 91.91\%, respectively. This work marks a significant improvement in attack detection capabilities. These studies demonstrate the advancement and efficacy of several algorithms in addressing the challenges connected with IoT botnet detection.

\section{Preliminaries}
The proposed model integrates 1D-CNN’s, BiLSTM and an attention mechanism to create a robust architecture for sequential data analysis. This hybrid approach capitalizes on CNN’s ability to detect local patterns and spatial relationships, BiLSTM’s capacity to capture long-range dependencies, and the attention mechanisms focus on crucial input elements. By incorporating two complimentary techniques, the model aims to boost interpretability and performance across a range of sequential data requirements, providing additional and sophisticated analysis than common DL strategies.
\subsection{1D-CNN’s}
One-Dimensional Convolutional Neural Network has shown particularly effective for classification methods, adopting a multi-layered design that consists of five different layers for processing.The Convolutional layers serve a substantial part in determining feature extraction and parameter optimization during back propagation. These structures make use of dedicated filters, incorporating pooling techniques to prevent over-fitting. The data is then turned into class representations using flattening and connected layers. The modular design provides structural flexibility to meet a variety of categorization requirements. The mathematical equation\cite{9781894} of the 1D – CNN are as follows:

\begin{equation}\label{1D-CNN}
   y_{i}^{l} = \sigma \left( \sum\limits_{i=1}^{N_{l-1}} \text{conv1D}(w_{i,j}^{l}, x_{i}^{l-1}) + b_{j}^{l} \right)
\end{equation}

\subsection{Bidirectional Long Short-Term Memory (Bi-LSTM)}
Bi-LSTM analyzes sequences bidirectionally to detect connections and temporal patterns. Its nature of two-way analysis improves its ability to recognize complicated sequential patterns.\par

\begin{figure}[h]
    \centering
    \includegraphics[width=0.9\linewidth]{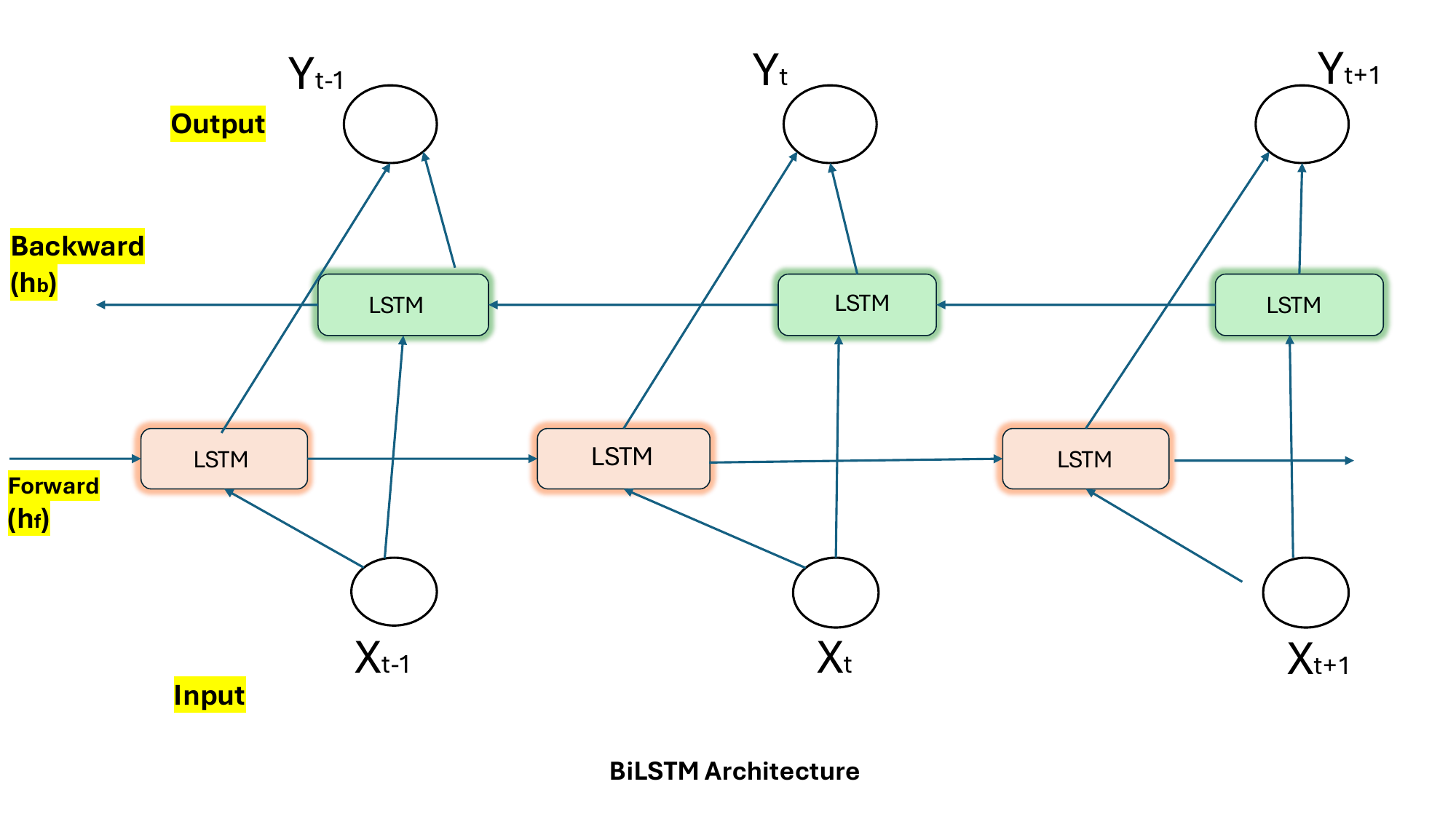}
    \caption{BiLSTM \cite{9696242}.}
    \label{fig1:BiLSTM}
\end{figure}
As this architecture employs two hidden layers: backward and forward (in Fig:1) The forward layers processes sequence from start to end, whereas the backward layer processes it in reverse way. After it combines the results from both layers as final output.Some important equations of that particular model are as follows:
\begin{equation}\label{BiLSTM}
    h_t^f = \tanh(W_{xh}^f x_t + W_{hh}^f h_{t-1}^f + b_h^f) 
\end{equation}
\begin{equation}\label{BiLSTM}
    h_t^b = \tanh(W_{xh}^b x_t + W_{hh}^b h_{t+1}^b + b_h^b) 
\end{equation}
\begin{equation}\label{BiLSTM}
    y_t = W_{hy}^f h_t^f + W_{hy}^b h_t^b + b_y
\end{equation}

\subsection{Attention Mechanism}
 The method utilizes dot product attention based on key, value and query connections.The key matrix, which originates from the value matrix, interacts with query vectors to present importance weights. These weights then lead process of aggregating value vectors.

\begin{equation}
\begin{aligned}
\mathbf{K} &= \tanh(\mathbf{V}\mathbf{W}^{a})
\end{aligned}
\label{eq:attention-K}
\end{equation}

\begin{equation}
\begin{aligned}
\mathbf{d} &= \operatorname{softmax}(\mathbf{q}\mathbf{K}^{T}), 
\quad \mathbf{a} = \mathbf{d}\mathbf{V}
\end{aligned}
\label{eq:attention-d}
\end{equation}

\begin{figure}[h]
    \centering
    \includegraphics[width=0.9\linewidth]{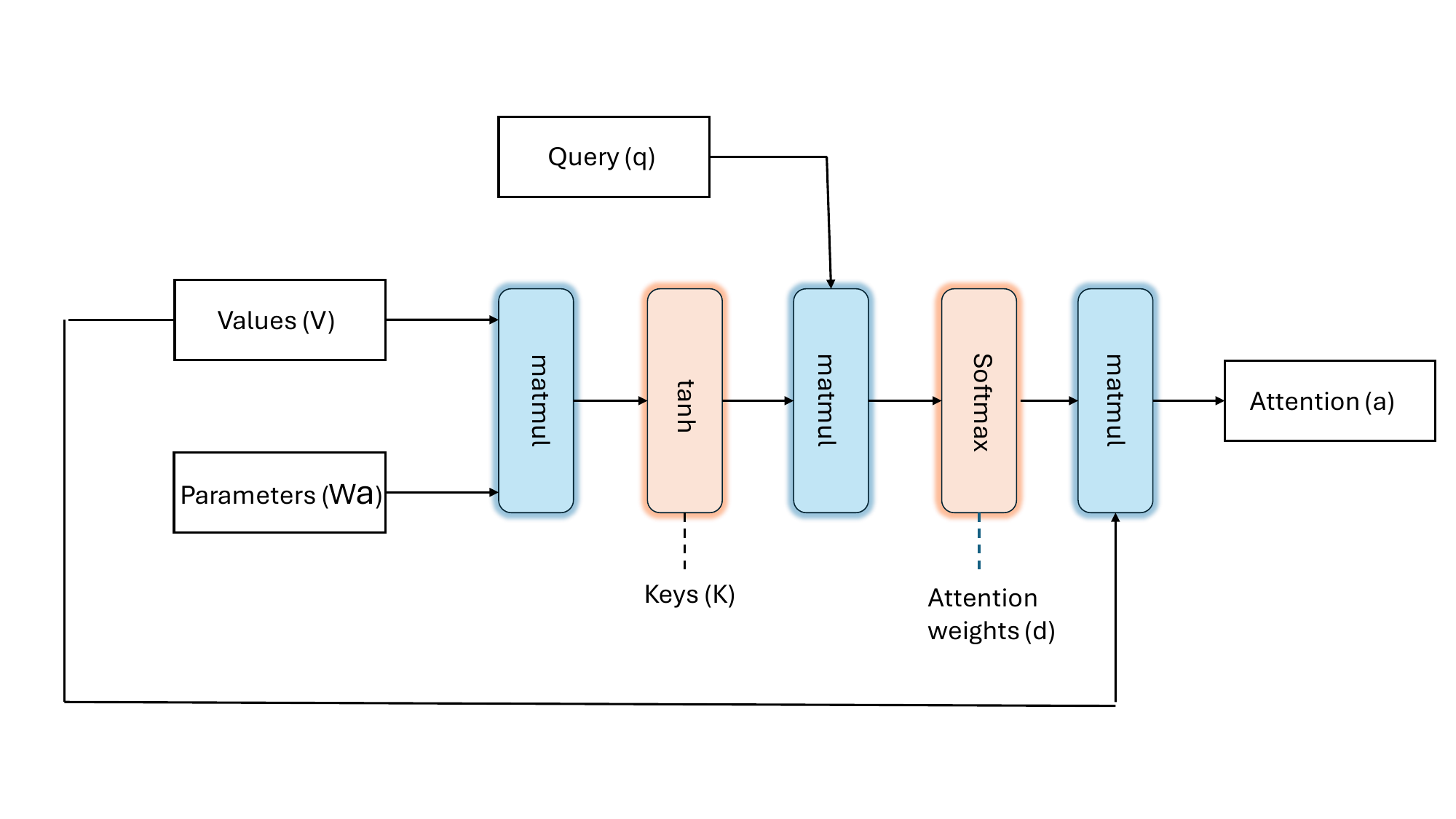}
    \caption{Attention Mechanism \cite{10.1145/3217871.3217872}.}
    \label{fig2:Attention Mechansim Architecture}
\end{figure}

\section{The Proposed Deep Learning Model}
\subsection{Hybrid model architecture}
In this work, we used N-BaIoT dataset, which consists of network traffic data from five categories of IoT devices: security cameras, webcams, doorbells,thermostats, baby monitors. This includes both ordinary traffic and traffic generated by several botnet assaults, namely Mirai and Bashlite (gafgyt). We implemented a model approach to correctly classify network data as normal or pertaining to one of nine different attack types. To improve performance this neural network architecture consists of three components one dimensional convolutional, bidirectional long short-term memory (Bi-LSTM), and attention process. The model's framework starts with a 1D convolutional with 128 filters, a kernel size of 5, and ReLU activation. This is preceded by batch normalization, maximum pooling and a 0.3\% dropout rate. Two more convolutional layers are added, one with 256 filters and the second is with 128 filters with a kernel size of three, maintaining an identical pattern. Temporal features are eventually examined by a Bi-LSTM layer of 128 units, which is supported by an attention mechanism. The structure continues with two dense layers of 256 and 128 units each, all with 0.4 rate of dropout and with ReLU activation. The final layer yields a 10 class softmax output. The optimization process Adam with learning rate of 0.001 is used in the model.
\begin{figure*}[!h]
    \centering
    \includegraphics[width=0.6\linewidth]{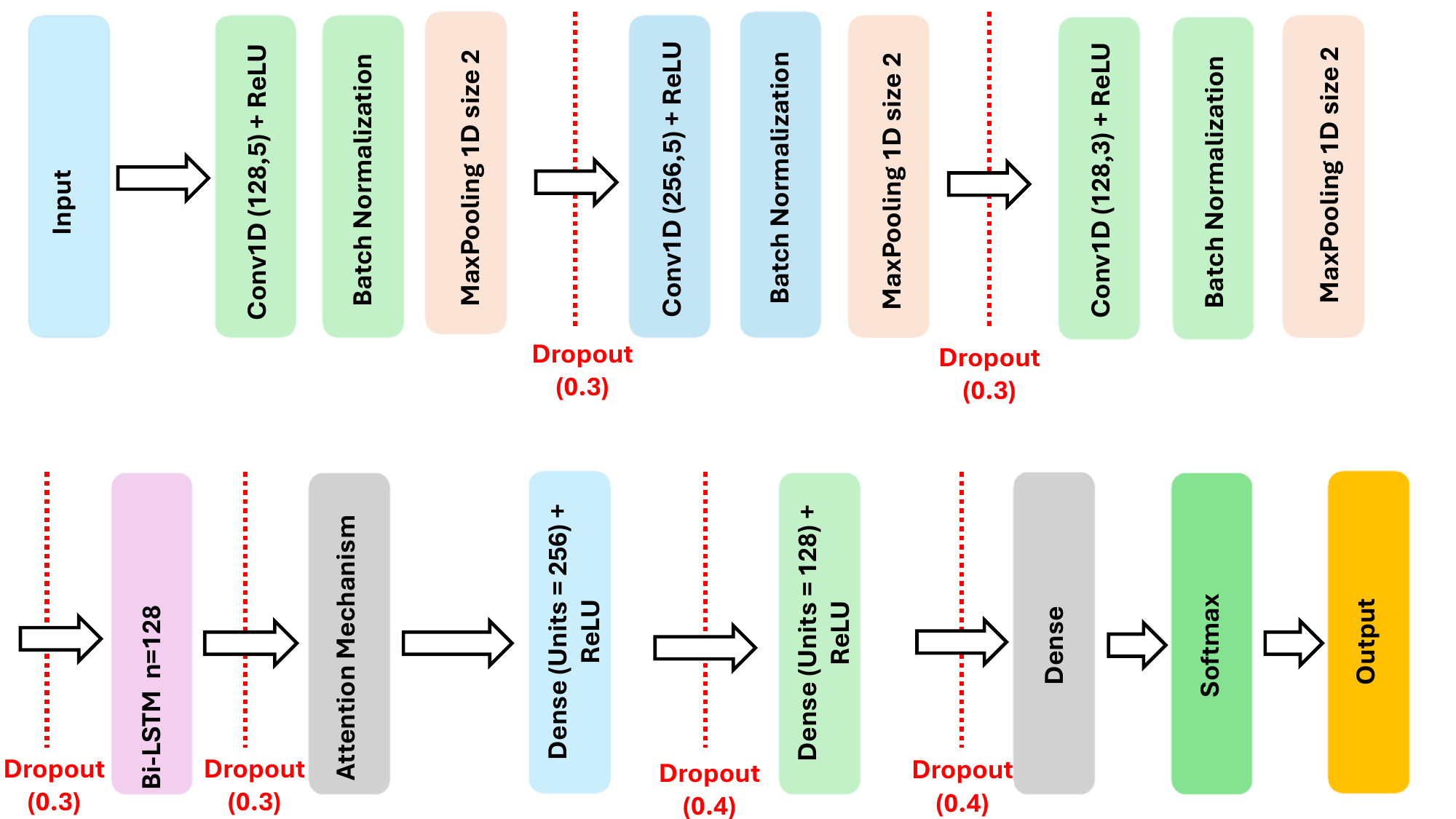}
    \caption{The Proposed Model Architecture}
    \label{fig3: proposed architecture}
\end{figure*}

\section{Experimental Setup}
To validate our model, we used the N-BaIoT dataset, allocating 20\% for testing and 80\% for training purposes. The model was trained on an NVIDIA T4 GPU, and the total training time was approximately 1 hour. During inference, the model achieves an average inference time of around 50 milliseconds per sample. The model’s training setup included 2 dropout layers (0.3, 0.4), and the Adam optimizer, used  sparse categorical cross-entropy  with softmax activation to provide multi-class classification output. 

\begin{table*}[!t]
\centering
\caption{Data Attack Schema}
\label{data_attack_schema}
\begin{tabular}{|l|l|r|r|r|r|r|r|}
\hline
\textbf{Botnet} & \textbf{Attack Type} & \textbf{Doorbell} & \textbf{Thermostat} & \textbf{Baby Monitor} & \textbf{Webcam} & \textbf{Security Camera 1} & \textbf{Security Camera 2} \\ \hline
\multicolumn{2}{|c|}{Benign}                     & 88648             & 13113               & 175240                & 52150            & 160688                   & 6613                      \\ \hline
{Bashlite} & COMBO                & 112732            & 53012               & 58152                 & 58669            & 118910                   & 113681                    \\ \cline{2-8} 
                         & JUNK                 & 58865             & 30312               & 28349                 & 28305            & 59966                    & 55992                     \\ \cline{2-8} 
                         & SCAN                 & 57969             & 27494               & 27859                 & 27698            & 57694                    & 56397                     \\ \cline{2-8} 
                         & TCP                  & 193677            & 95021               & 92581                 & 97783            & 193897                   & 186891                    \\ \cline{2-8} 
                         & UDP                  & 209807            & 104791              & 105782                & 110617           & 208669                   & 206700                    \\ \hline
 {Mirai}    & ACK                  & 102195            & 113285              & 91123                 & 0                & 118551                   & 218667                    \\ \cline{2-8} 
                         & SCAN                 & 107685            & 43192               & 103621                & 0                & 193877                   & 89604                     \\ \cline{2-8} 
                         & SYN                  & 122573            & 116807              & 118128                & 0                & 127597                   & 248194                    \\ \cline{2-8} 
                         & UDP                  & 237665            & 151481              & 217034                & 0                & 314856                   & 308963                    \\ \cline{2-8} 
                         & Plain UDP            & 81982             & 87368               & 80808                 & 0                & 110466                   & 162680                    \\ \hline
\textbf{Total}            &                      & 1373798           & 835876              & 1098677               & 375222           & 1665151                  & 1713882                   \\ 
\hline
\end{tabular}
\end{table*}

\subsection{Dataset}
The N-BaIoT dataset\cite{nbaiot_dataset} incorporates 7,062,606 network traffic data patterns from several smart devices to identify attacks like botnet. It comprises of 115 features determined across many time frames, involving both valid and malicious activities on the network. It includes traffic data from nine business IoT devices infected with malware either Mirai or BASHLITE,conveying 10 various attack types. This rich dataset provides a robust foundation for developing and evaluating intrusion detection systems in IoT networks. The dataset has 10 classes of attacks with 1 class as benign.

Table 1. The above illustrates the schema and the number of rows that are present to apply our models.
\subsubsection{Data Gathering and Preprocessing}
\begin{itemize}
    \item Data Gathering: Normal traffic data was collected immediately after the installation of the IoT devices. Subsequently, the devices were infected with different botnet attacks, and the resulting traffic was captured using Wireshark.
    \item Feature Extraction: From the captured traffic, 115 features were extracted, encompassing statistical information over various time intervals ranging from 100 milliseconds to one minute.
    \item Normalization: Given that the dataset consists solely of numeric features, we applied the Min-Max scaler to normalize the data. This scaling technique ensures that all features are within the range [0, 1], facilitating the training process and improving model performance.
    \item Class Balancing: To address class imbalance, we ensured that each class (benign and the 9 types of botnet attacks out of 10) had an equal number of samples. Specifically, we used 10,000 samples per class, resulting in a balanced dataset that mitigates bias and enhances the robustness of our model.
\end{itemize}

\subsection{Evaluation Parameters}
\subsubsection{Performance Parameters}
To analyze the performance of the architecture, we use standard metrics resulting from the confusion matrix. These indicators comprise of accuracy, recall, precision, and F1-score all these provides a dept insight into the model's classification capabilities. Four key elements are involved in confusion matrix:
\begin{enumerate}
 \item   False Positives (FP): Normal instances mistakenly classified as attacks.
    \item 	False Negatives (FN): Attack instances mistakenly classified as normal.
    \item   True Positives (TP): Correctly identified attack instances.
    \item   True Negatives (TN): Correctly identified normal instances.

\end{enumerate}

 Using these elements, we calculate the following indicators:
1. \textbf{\textit{Accuracy:}} The percentage of accurate predictions (which involves true positives and true negatives) across all instances examined.
\begin{equation}\label{Accuracy}
   \text{Accuracy} = \frac{TP + TN}{TP + FP + TN + FN}
\end{equation}

2. \textbf{\textit{Recall:}} The output of accurately determined positive instances to the total number of positive instances.

\begin{equation}\label{Recall}
   \text{Recall} = \frac{TP}{TP + FN}
\end{equation}

3. \textbf{\textit{Precision:}} The proportion of precisely detected positive cases to all anticipated positive instances.

\begin{equation}\label{Precision}
   \text{Precision} = \frac{TP}{TP + FP}
\end{equation}

4. \textbf{\textit{F1- Score:}} The harmonic mean of recall as well as precision, presents a fair evaluation of the performance of respective model.

\begin{equation}\label{F1-Score}
   F1\text{-Score} = 2 \times \frac{Precision \times Recall}{Precision + Recall}
\end{equation}

\subsection{Reliability Parameters}
The evaluation procedure includes two advanced sophisticated statistical tools: Kappa coefficient and Matthew’s correlation. These additional measures improve the procedure by analyzing relationships between classes and considering factors of random agreement, providing a more thorough insight  of classifier performance. \par
\subsubsection{Cohen’s Kappa Co-efficient}
It is a statistic used for evaluating classification ability that examines the concordance between actual and anticipated results while incorporating agreements that may arise by chance \cite{yilmaz2023weighted}. This coefficient operates on a scale of negative one to positive one, with positive one indicating ideal agreement, zero indicating chance acceptance and negative one signifying contradiction. This technique provides a more detailed evaluation than traditional measures.\par
\subsubsection{MCC}
It provides a sophisticated measure of classification quality, particularly useful for datasets with unbalanced class distribution \cite{chicco2023matthews}. This complete measure uses all elements of confusion matrix – TP, FP,TN, and FN, to calculate a score ranging from negative to positive one. A flawless anticipated accuracy is represented by a score of +1, 0 yielding a random guessing score, and -1 represents entirely opposite predictions. As a result, MCC is especially useful for evaluating classifier performance when working with unbalanced datasets.
\section{Results and Analysis}
\subsection{Training Results}
We appraise the results of the hybrid approach on the basis of recall, accuracy, precision, F1- score and correlation coefficients. The training results are shown in Figure 5.

\begin{figure}[h]
    \centering
    \includegraphics[width=0.9\linewidth]{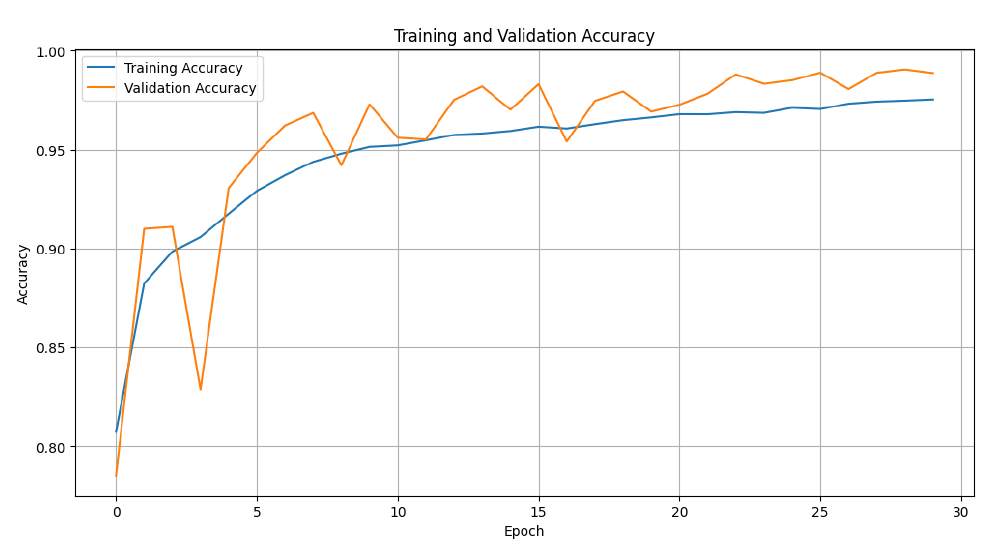}
    \caption{Accuracy Curve}
    \label{fig4:Accuracy}
\end{figure}

\begin{figure}[h]
    \centering
    \includegraphics[width=0.9\linewidth]{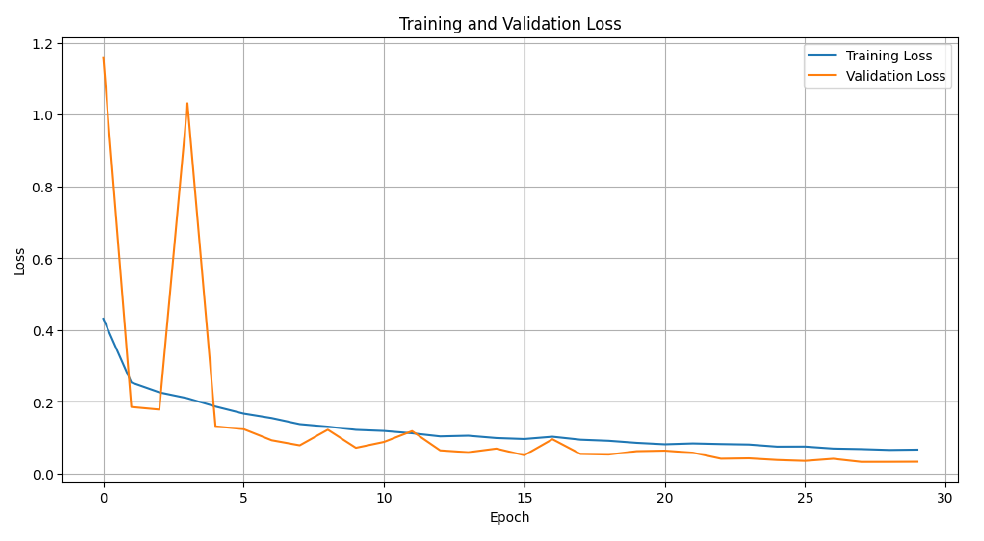}
    \caption{Loss Curve}
    \label{fig5:Loss}
\end{figure}

\begin{itemize}
    \item Loss: The decreasing trend in the loss curves implies an ongoing reduction in the error rate of the architecture on the training dataset as epochs rise up. This pattern indicates that the model is actively improving and learning its predictive skills.
    \item Accuracy: The accuracy curves rise correspondingly, showing that the proposed model's predictions are growing accurately.
    \item Stability Near 30\textsuperscript{th} Epoch: The model’s accuracy stabilizes around the 30\textsuperscript{th} Epoch, suggesting it has learned, effectively identify underlying trends in the data and beyond that it will start to overfit.
\end{itemize}
\subsection{Confusion Matrix and Classification Report}

\begin{figure}[ht]
    \centering
    \includegraphics[width=0.9\linewidth]{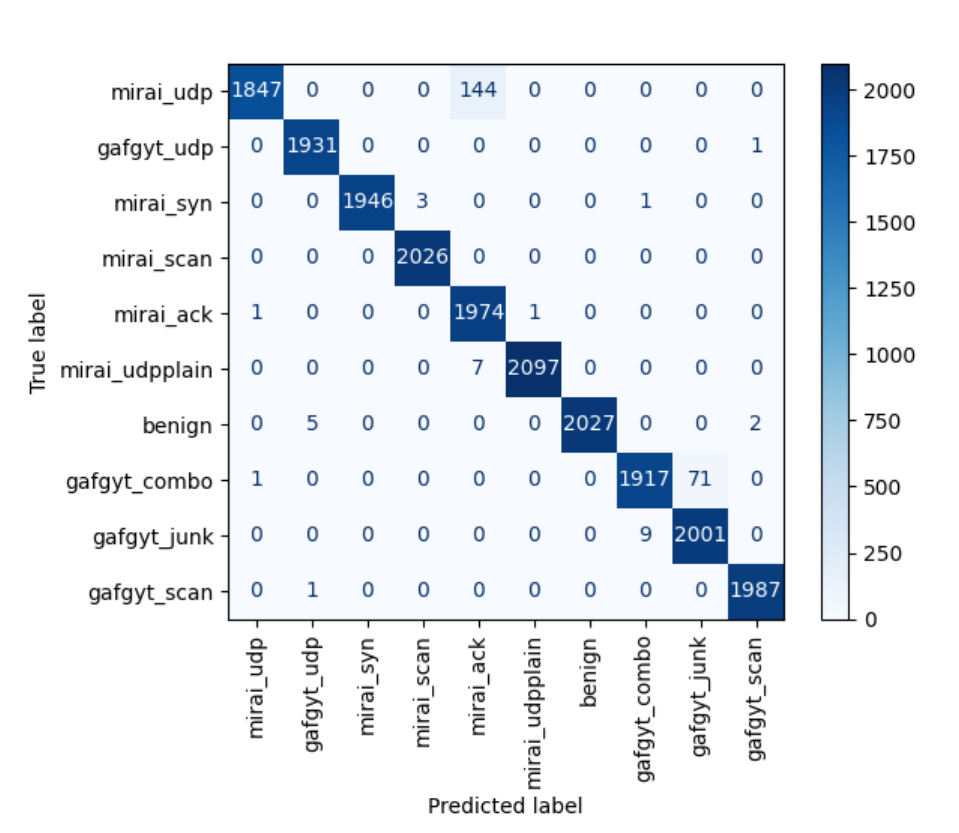}
    \caption{Confusion Matrix}
    \label{fig6:Confusion}
\end{figure}

The confusion matrix is utilized to evaluate the classifiers performance comprehensively. When most of the samples in the matrix are concentrated along the diagonal, as illustrated in Figure 5, the model can accurately differentiate between attack samples and normal samples. 
\begin{itemize}
    \item \textbf{Class-Specific Metrics:}
    \item \texttt{mirai\_udp}: High precision (1.00) but lower recall (0.93), suggesting the model is very accurate when it predicts this class, but it misses some instances.
    \item \texttt{mirai\_ack}: Lower precision (0.93) but perfect recall (1.00), meaning the model identifies all instances but has some false positives.
    \item \texttt{Remaining classes }: High scores across all metrics, indicating strong performance with minor variations.
\end{itemize}

\begin{table}[h]
\centering
\caption{Classification report of each class}
\begin{tabular}
{|c|p{0.8cm}|p{0.8cm}|p{0.8cm}|p{0.8cm}|p{0.8cm}|p{0.8cm}|}
\hline
\textbf{Class} & \textbf{Precision} & \textbf{Recall} & \textbf{F1-Score} & \textbf{Support} \ & \textbf{Accuracy} \\ \hline
benign & 1.00 & 0.996 & 0.998 & 2034 & 0.997 \\ \hline
gafgyt\_combo & 0.995 & 0.964 & 0.979 & 1989 & 0.964 \\ \hline
gafgyt\_junk & 0.966 & 0.995 & 0.980 & 2010 & 0.996\\ \hline
gafgyt\_scan & 0.998 & 0.999 & 0.999 & 1988 & 0.999 \\ \hline
gafgyt\_udp & 0.997 & 0.999 & 0.998 & 1932 & 0.928 \\ \hline
mirai\_ack & 0.929 & 0.999 & 0.963 & 1976 & 0.999 \\ \hline
mirai\_scan & 0.998 & 1.00 & 0.999 & 2026 & 1.00 \\ \hline
mirai\_syn & 1.00 & 0.999 & 0.999 & 1950 & 0.998 \\ \hline
mirai\_udp & 0.999 & 0.928 & 0.962 & 1991 & 0.928\\ \hline
mirai\_udpplain & 0.999 & 0.997 & 0.998 & 2104 & 0.997 \\ \hline
accuracy & 0.988 & 0.988 & 0.988 & 0.987 &  \\ \hline
Macro-Average & 0.988 & 0.988 & 0.988 & 20000 &  \\
\hline
Weighted-Average & 0.988 & 0.988 & 0.988 & 20000 &  \\
\hline

\end{tabular}
\end{table}

\subsection{Reliability Parameters}
\subsubsection{Kappa Coefficient}
It assesses the agreement between actual and anticipated classifications after controlling for chance agreement. A kappa score of 0.985\% indicates practically perfect agreement
\subsubsection{Mathews Correlation Coefficient}
It measures the level of binary classifications. It suggests 0.986\% a high connection between expected and true classifications.
\begin{table}[htbp]
\centering
\caption{Cohen's Kappa and Matthews Correlation Coefficient (MCC) for each class}
\begin{tabular}{|c|c|c|}
\hline
\textbf{Class} & \textbf{Cohen's Kappa} & \textbf{Matthews} \\ \hline
benign & 0.998 & 0.998 \\ \hline
gafgyt\_combo & 0.976 & 0.976 \\ \hline
gafgyt\_junk & 0.978 & 0.978 \\ \hline
gafgyt\_scan & 0.998 & 0.998 \\ \hline
gafgyt\_udp & 0.997 & 0.997 \\ \hline
mirai\_ack & 0.958 & 0.959 \\ \hline
mirai\_scan & 0.999 & 0.999 \\ \hline
mirai\_syn & 0.999 & 0.999 \\ \hline
mirai\_udp & 0.957 & 0.959 \\ \hline
mirai\_udpplain & 0.997 & 0.997 \\ \hline
\end{tabular}
\end{table}

\subsection{ROC Curve}
The Receiver Operating Characteristics curve demonstrates the connection among sensitivity at several thresholds. As illustrated in fig:6, the curve for the hybrid approach indicating great performance. This suggests that model generates a large number of true positives while maintaining a low rate of false positives. These results presented that the respective hybrid approach is effective in distinguishing between classes like negative and positive.

\begin{figure}[h]
    \centering
    \includegraphics[width=0.9\linewidth]{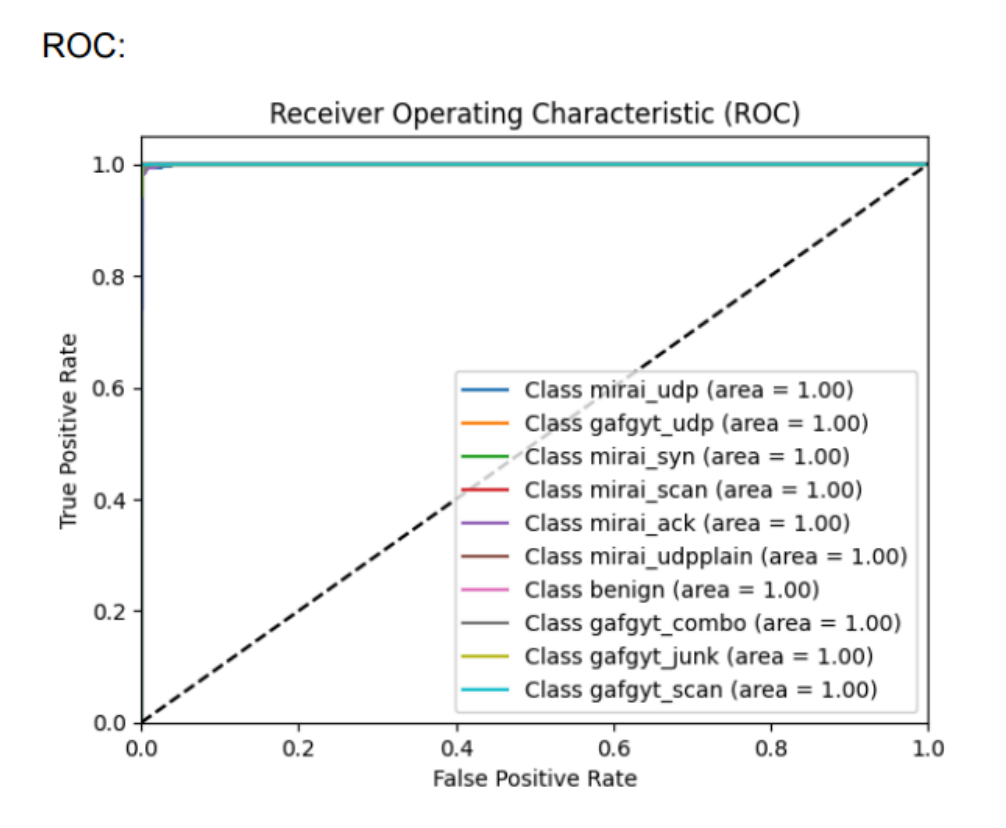}
    \caption{ROC Curve}
    \label{fig7:ROC}
\end{figure}

\section{Comparative Analysis with Similar Architectures}

\begin{table*}[htbp]
\centering
\caption{Comparative Analysis}
\begin{tabular}{|c|p{0.9cm}|p{0.9cm}|p{0.9cm}|p{0.9cm}|p{0.9cm}|p{0.9cm}|p{2.5cm}|p{2.5cm}|p{2.5cm}|}
\hline
\textbf{Class} & \textbf{1D-CNN Precision\cite{majd2023iot}} & \textbf{1D-CNN Recall\cite{majd2023iot}} & \textbf{1D-CNN F1-Score\cite{majd2023iot}} & \textbf{CNN-LSTM Precision\cite{majd2023iot}} & \textbf{CNN-LSTM Recall\cite{majd2023iot}} & \textbf{CNN-LSTM F1-Score \cite{majd2023iot}} & \textbf{1D-CNN-BiLSTM Attention mechanism Precision} & \textbf{1D-CNN-BiLSTM Attention mechanism Recall} & \textbf{1D-CNN-BiLSTM Attention mechanism F1-Score} \\ 
\hline
Benign & 1 & 1 & 1 & 1 & 1 & 1 & 1 & 0.99 & 0.99 \\ 
\hline
 gafgyt\_combo & 0.93 & 0.84 & 0.88 & 0.93 & 0.88 & 0.90 & 1 & 0.96 & 0.98 \\ 
\hline
 gafgyt\_junk & 0.81 & 0.94 & 0.87 & 0.85 & 0.94 & 0.89 & 0.97 & 1 & 0.98 \\ 
\hline
 gafgyt\_scan & 1 & 1 & 1 & 1 & 1 & 1 & 1 & 1 & 1 \\ 
\hline
 gafgyt\_udp & 0.5 & 1 & 0.66 & 0.55 & 1 & 0.66 & 1 & 1 & 1 \\ 
\hline
 mirai\_ack & 0.99 & 0.99 & 0.99 & 1 & 1 & 1 & 0.93 & 1 & 0.96 \\ 
\hline
 mirai\_scan & 0.98 & 1 & 0.99 & 0.99 & 1 & 0.99 & 1 & 1 & 1 \\ 
\hline
 mirai\_syn & 1 & 0.99 & 0.99 & 1 & 0.99 & 0.99 & 1 & 1 & 1 \\ 
\hline
 mirai\_udp & 0.96 & 0.91 & 0.93 & 1 & 0.93 & 0.96 & 1 & 0.93 & 0.96 \\ 
\hline
mirai\_udpplain & 0.91 & 0.96 & 0.93 & 0.93 & 0.97 & 0.95 & 1 & 1 & 1 \\ 
\hline
\end{tabular}
\end{table*}

Table IV, demonstrates a contrast between \cite{majd2023iot} and our model in terms of evaluation metrics. In Table IV, the proposed hybrid deep learning strategy defeated all other models, obtaining superior accuracy than other approaches. Higher Precision and F1-Scores from Comparison Table: Our proposed approach deliberately achieves greater precision and F1-scores among several classes types, demonstrating well accuracy and balance. Perfect Scores in Critical Classes: For critical classes like Bashlite UDP and Mirai UDPplain, the proposed model achieves perfect scores, demonstrating superior performance. Balanced Performance: Even in classes where our model doesn’t achieve perfect scores, it maintains a good balance between precision and recall, obtaining a large output of  F1-scores. Overall, the 1D-CNN+BiLSTM+Attention model demonstrates superior performance, particularly in achieving higher precision and F1-scores, making it more reliable and effective for the given classification task.

\section{Conclusion}
This paper proposed and evaluated a deep learning model for botnet attack detection, which incorporates an attention process into a hybrid 1D- convolutional and Bi-LSTM architecture. The proposed model demonstrated 99\% accuracy and close to ideal results for reliability parameters including Matthews Correlation coefficient and Cohen's kappa coefficient. Moreover, the optimized architecture ensured minimal computation costs, making it perfect for real-time monitoring and protection of IoT devices. 

\bibliographystyle{IEEEtran}
\bibliography{main}

@ARTICLE{9520818,
  author={Li, Yuxi and Zuo, Yue and Song, Houbing and Lv, Zhihan},
  journal={IEEE Internet of Things Journal}, 
  title={Deep Learning in Security of Internet of Things}, 
  year={2022},
  volume={9},
  number={22},
  pages={22133-22146},
  doi={10.1109/JIOT.2021.3106898}}

@INPROCEEDINGS{9120761,
  author={Khoa, Tran Viet and Saputra, Yuris Mulya and Hoang, Dinh Thai and Trung, Nguyen Linh and Nguyen, Diep and Ha, Nguyen Viet and Dutkiewicz, Eryk},
  booktitle={2020 IEEE Wireless Communications and Networking Conference (WCNC)}, 
  title={Collaborative Learning Model for Cyberattack Detection Systems in IoT Industry 4.0}, 
  year={2020},
  volume={},
  number={},
  pages={1-6},
  doi={10.1109/WCNC45663.2020.9120761}}

@INPROCEEDINGS{10136065,
  author={Sakthipriya, N and Govindasamy, V. and Akila, V.},
  booktitle={2023 2nd International Conference on Paradigm Shifts in Communications Embedded Systems, Machine Learning and Signal Processing (PCEMS)}, 
  title={A Comparative Analysis of various Dimensionality Reduction Techniques on N-BaIoT Dataset for IoT Botnet Detection}, 
  year={2023},
  volume={},
  number={},
  pages={1-6},
  doi={10.1109/PCEMS58491.2023.10136065}}

@article{https://doi.org/10.1155/2021/3806459,
author = {Alkahtani, Hasan and Aldhyani, Theyazn H. H.},
title = {Botnet Attack Detection by Using CNN-LSTM Model for Internet of Things Applications},
journal = {Security and Communication Networks},
volume = {2021},
number = {1},
pages = {3806459},
doi = {https://doi.org/10.1155/2021/3806459},
url = {https://onlinelibrary.wiley.com/doi/abs/10.1155/2021/3806459},
eprint = {https://onlinelibrary.wiley.com/doi/pdf/10.1155/2021/3806459},
year = {2021}
}

@INPROCEEDINGS{9781894,
  author={Ibrahim, Moamen and Badran, Khaled M. and Esmat Hussien, Ahmed},
  booktitle={2022 13th International Conference on Electrical Engineering (ICEENG)}, 
  title={Artificial intelligence-based approach for Univariate time-series Anomaly detection using Hybrid CNN-BiLSTM Model}, 
  year={2022},
  volume={},
  number={},
  pages={129-133},
  doi={10.1109/ICEENG49683.2022.9781894}}

@ARTICLE{9696242,
  author={Yousaf, Kanwal and Nawaz, Tabassam},
  journal={IEEE Access}, 
  title={A Deep Learning-Based Approach for Inappropriate Content Detection and Classification of YouTube Videos}, 
  year={2022},
  volume={10},
  number={},
  pages={16283-16298},
  doi={10.1109/ACCESS.2022.3147519}}

@inproceedings{10.1145/3217871.3217872,
author = {Brown, Andy and Tuor, Aaron and Hutchinson, Brian and Nichols, Nicole},
title = {Recurrent Neural Network Attention Mechanisms for Interpretable System Log Anomaly Detection},
year = {2018},
isbn = {9781450358651},
publisher = {Association for Computing Machinery},
address = {New York, NY, USA},
url = {https://doi.org/10.1145/3217871.3217872},
doi = {10.1145/3217871.3217872},
booktitle = {Proceedings of the First Workshop on Machine Learning for Computing Systems},
articleno = {1},
numpages = {8},
location = {Tempe, AZ, USA},
series = {MLCS'18}
}

@misc{nbaiot_dataset,
  author       = {Kashif Nazir},
  title        = {N-BaIoT Dataset},
  year         = {2020},
  url          = {https://www.kaggle.com/datasets/mkashifn/nbaiot-dataset},
  note         = {Accessed: 2024-10-26}
}

@article{yilmaz2023weighted,
  title={Weighted kappa measures for ordinal multi-class classification performance},
  author={Yilmaz, Ayfer Ezgi and Demirhan, Haydar},
  journal={Applied Soft Computing},
  volume={134},
  pages={110020},
  year={2023},
  publisher={Elsevier}
}

@article{chicco2023matthews,
  title={The Matthews correlation coefficient (MCC) should replace the ROC AUC as the standard metric for assessing binary classification},
  author={Chicco, Davide and Jurman, Giuseppe},
  journal={BioData Mining},
  volume={16},
  number={1},
  pages={4},
  year={2023},
  publisher={Springer}
}

@INPROCEEDINGS{10471929,
  author={Do, Phuc Hao and Le, Tran Duc and Vishnevsky, Vladimir and Berezkin, Aleksandr and Kirichek, Ruslan},
  booktitle={2024 26th International Conference on Advanced Communications Technology (ICACT)}, 
  title={A Horizontal Federated Learning Approach to IoT Malware Traffic Detection: An Empirical Evaluation with N-BaIoT Dataset}, 
  year={2024},
  volume={},
  number={},
  pages={1494-1506},
  doi={10.23919/ICACT60172.2024.10471929}}

@article{popoola2020hybrid,
  title={Hybrid deep learning for botnet attack detection in the internet-of-things networks},
  author={Popoola, Segun I and Adebisi, Bamidele and Hammoudeh, Mohammad and Gui, Guan and Gacanin, Haris},
  journal={IEEE Internet of Things Journal},
  volume={8},
  number={6},
  pages={4944--4956},
  year={2020},
  publisher={IEEE}
}

@INPROCEEDINGS{10836622,
  author={Hossain, Mahnaz and Al Mamun Rudro, Rifat and Razzaque, Rupom and Nur, Kamruddin},
  booktitle={2024 International Conference on Decision Aid Sciences and Applications (DASA)}, 
  title={Machine Learning Approaches for Detecting IoT Botnet Attacks: A Comparative Study of N-BaIoT Dataset}, 
  year={2024},
  volume={},
  number={},
  pages={1-7},
  doi={10.1109/DASA63652.2024.10836622}}

@inproceedings{majd2023iot,
  title={IoT Botnet Classification using CNN-based Deep Learning},
  author={Majd, Nahid Ebrahimi and Gudipelly, Dhatri Sai Kumar Reddy},
  booktitle={2023 IEEE International Performance, Computing, and Communications Conference (IPCCC)},
  pages={46--51},
  year={2023},
  organization={IEEE}
}

\end{document}